\documentclass[]{aa}
\usepackage{graphicx}
\usepackage{txfonts}
\begin{document}

\title{The pre-main sequence spectroscopic binary UZ Tau East: 
improved orbital parameters and accretion phase dependence 
\thanks{Based on observations made with the Isaac Newton 
Telescope, operated on the island of La Palma by the Isaac Newton
Group in the Spanish Observatorio del Roque de los Muchachos of the
Instituto de Astro{f\'\i}sica de Canarias, the ESO 3.6-meter telescope at 
La Silla Observatory in Chile, and the Shane 
3-meter telescope at Lick observatory in California.} }

\author{Eduardo L. Mart\'\i n\inst{1,2}, Antonio Magazz\`u\inst{3}\thanks{On leave from 
INAF-Osservatorio Astrofisico di Catania},  
Xavier Delfosse\inst{4}, Robert D. Mathieu\inst{5}}

\offprints{Eduardo L. Mart\'\i n}

\institute {Instituto de Astrof\'{\i}sica de Canarias, 38200 La Laguna, Tenerife, Spain \\ 
\email{ege@iac.es}
\and University of Central Florida, Department of Physics, PO Box 162385, Orlando, FL 32816-2385, 
USA \\
\email{ege@physics.ucf.edu}
\and  INAF-Telescopio Nazionale Galileo, Apartado 565, E-38700, Santa Cruz de La Palma, Spain \\ 
\email{magazzu@tng.iac.es}
\and Laboratoire d'Astrophysique Observatoire de Grenoble, B.P. 53, 414 rue de la Piscine, F-38041, Grenoble Cedex 9, France \\
\email{delfosse@obs.ujf-grenoble.fr}
\and Department of Astronomy, University of Wisconsin-Madison, 
475 North Charter Street, Madison, WI 53706, USA \\  
\email{mathieu@astro.wisc.edu}}

\date{Received /Accepted} 

\abstract 
{We present radial-velocity measurements obtained using 
high- and intermediate-resolution spectroscopic observations of the 
classical T Tauri star  UZ Tau East obtained from 1994 to 1996. 
We also provide  measurements of  
H$\alpha$ equivalent widths and optical veiling. 
Combining our radial-velocity data with those  
recently reported by Prato et al. (2002), we improve the orbital elements for 
this spectroscopic binary. The orbital period is 18.979$\pm$0.007 days and the eccentricity is 
e=0.14.  We find variability in the H$_\alpha$ emission and veiling, signposts 
of accretion, but at periastron passage the accretion is not as clearly 
enhanced as in the case of the binary DQ Tau. The difference in the behaviour of these two 
binaries is consistent with the hydrodynamical models of accretion from circumbinary disks   
 because UZ Tau East has lower eccentricity than DQ Tau. It seems that enhanced periastron 
accretion may occur only in systems with very high eccentricity (e$>$0.5).

 }

\authorrunning{Mart{\'\i}n et al.}
\titlerunning{The spectroscopic binary UZ Tau East}

\markboth{The spectroscopic binary UZ Tau East}{Mart{\'\i}n et al.}

\maketitle

\section{Introduction}

UZ Tau was one of the first 11 stars originally identified as members of the class of T Tauri 
variable stars, and it was also noticed to be a wide pair with angular separation 
3\farcs6 (Joy \& van Biesbroek 1944). UZ Tau East and West are a pair of 
classical T Tauri stars (CTTSs), i.e. very young stars with an emission line spectrum 
that indicates active mass accretion from a circumstellar disk
(Appenzeller \& Mundt 1989; Bertout 1989). 

UZ Tau E is a CTTS with spectral type M1, strong H$_\alpha$ emission,   
strong excess continuum emission at infrared and submillimiter wavelengths 
(Jensen, Mathieu \& Fuller 1996) and a Keplerian disk (Simon et al. 2000). 
UZ Tau W is itself a pair with an angular separation of 
0\farcs34 (47.6 AU at 140 pc; Simon et al. 1995). 
UZ Tau W does not have as strong an infrared excess as E, and it also has much less 
submillimiter emission (Jensen et al. 1996). 

Prior to our observations, UZ Tau E was thought to be a single star. 
In 1994 we started high-resolution spectroscopic observations of this star and 
we noticed the radial-velocity variability. Follow-up observations were obtained 
in several runs.  
An early report of the orbital parameters of UZ Tau E 
 was given by Mathieu, Mart\'{\i}n  \& Magazz\`u (1996). Prato et al. (2002) obtained 
radial-velocity measurements of both components of UZ Tau E using near-infrared spectra, 
and estimated the mass ratio of the system. 

After AK Sco (Andersen et al. 1989), GW Ori (Mathieu et al. 1991), 
DQ Tau (Mathieu et al. 1997), V4046 Sgr 
(Quast et al. 2000), and RX J0530.7-0434 (Covino et al. 2001), 
UZ Tau E is only the fifth CTTS known to be a spectroscopic binary. 
Another CTTS spectroscopic binary, namely KH 15D, has recently been 
reported by Johnson et al. (2004). 

In this paper we present our spectroscopic observations of UZ Tau E 
obtained from 1994 to 1996. 
In Sect. 2 we describe the observations. Section 3 contains the main results, including 
the orbital solution found using our data and those in the literature. 
Section 4 provides a discussion of the properties of this spectroscopic binary.

\section{Observations and Data Reduction.}

Most of the spectroscopic observations presented in this paper 
were carried out with the 2.5-meter Isaac Newton 
Telescope (INT) equipped with the Intermediate Dispersion Spectrograph (IDS). 
We used the cameras and gratings listed in Table 1 which provided the full width half maximum 
(FWHM) spectral resolutions given in 
the same table. We placed both UZ Tau E and W along the same IDS long slit.  

The initial identification of radial-velocity variations was made in the analysis 
of the IDS spectra. In order to confirm the results and to improve the orbital parameters, 
we supplemented the INT data with observations obtained with the Hamilton echelle at 
the Shane 3-meter telescope at Lick Observatory. We measured the radial velocity 
of UZ Tau E in a spectrum obtained in 1988, which has been used to derive veiling by 
Basri \& Batalha (1990) and to measure the strength of the lithium resonance line 
by Basri, Mart\'{\i}n  \& Bertout (1991). Additional observations were obtained at 
the ESO 3.6-meter telescope in January 1996, using the CASPEC echelle spectrograph, 
and again at Lick with the Hamilton spectrograph in March 1996. Table 1 gives the log 
of the spectroscopic observations.  

IDS data reduction was done with IRAF\footnote{The IRAF package is distributed by the National 
Optical Astronomical Observatories.} routines in the same manner as described 
in Mart\'{\i}n  et al. (1992). Hamilton data reduction was made with IDL routines 
as explained in Basri \& Batalha (1990). CASPEC data reduction was carried out 
with the echelle package in IRAF. The data were unbiased and flatfielded before extracting 
the spectrum. Wavelength calibration was performed using a ThAr lamp spectrum obtained the same 
night. 


\section{Results.}

Using the IDS spectra we measured radial velocities for UZ Tau E by cross-correlation 
with the spectra of UZ Tau W observed simultaneously (the slit was aligned with the visual 
binary axis). After several tests, we found that the spectral region between 640~nm and 650~nm 
gave the most precise radial velocities. Figure 1 displays one example of the spectral 
region that we used for cross-correlating the spectra of UZ Tau E and W. 
Table 2 lists the radial velocities measured in our spectra.  These velocities are heliocentric, 
and relative to an assumed heliocentric radial velocity for UZ Tau W of 19.3 km s$^{-1}$.  
The radial-velocity error bars were obtained from the cross-correlation fit given by the 
IRAF task fxcor. 

   \begin{figure*}
   \centering
   \includegraphics[width=\textwidth]{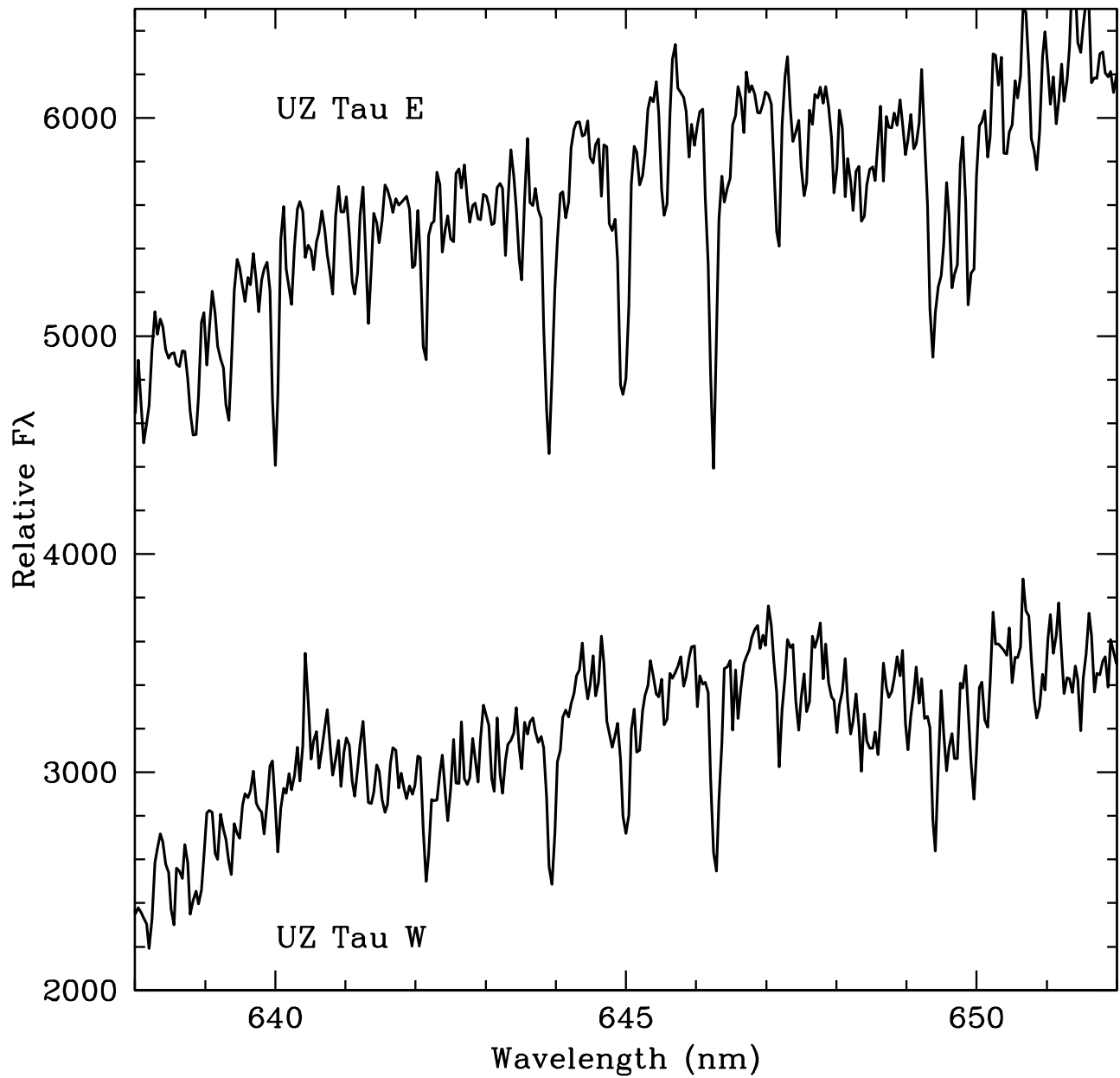}
   \caption{Zoom on the IDS spectral region that we used for measuring radial velocities. 
These IDS spectra of UZ Tau E and W were obtained on 20 Nov 1995.}
   \end{figure*}

We also measured H$\alpha$ equivalent widths and optical veiling in the IDS spectra, 
and we provide the values in Table 2. 
We did not measure these quantities in the echelle spectra (from CASPEC and Hamilton) 
because their resolution is much higher than that of IDS. It is well know that equivalent 
width measurements are affected by the spectral resolution. 
Moreover, we only have a few 
echelle spectra, and hence they do not contribute significantly to the study of the 
variability of H$_\alpha$ emission and veiling as a function of orbital phase. 
Nevertheless, in Table 2 we do quote the veiling measured by Basri \& Batalha (1989). 
 Our H$\alpha$ equivalent widths 
are estimated to be accurate within 10\% at a 3~$\sigma$ confidence level.

The veiling at a given wavelength (denoted as v$_{\lambda}$) is 
 defined as the ratio of the excess flux to the photospheric flux. As a consequence, 
we have measured veiling as the ratio of two equivalent widths, i.e. 
v$_{\lambda}$=EW(LiI)$_{\rm phot}$/EW(LiI)$_{\rm obs}$ - 1, where EW(LiI)$_{\rm obs}$ 
is the observed equivalent width of the lithium doublet at 6708 \AA . 
We could not obtain a value of veiling for the spectrum at JD=49972.730 using this 
method,  
because the spectral range of it did not include the lithium feature. Thus, the value 
given in Table~2 was obtained from the same spectral region as was used for the radial 
velocity measurements. We artificially veiled the spectrum of UZ Tau E obtained at 
JD=50040.528 by a small amount until we found a good match.
We used  EW(LiI)$_{\rm phot}$=720 m\AA , obtained by Basri et al. (1991). This value 
is close to the photospheric lithium equivalent width  
typical of early M-type T Tauri stars  
(Mart\'{\i}n  1997). While the absolute veiling depends on this 
assumption, we note that our main interest is to measure the 
variability of the relative veiling as a function of orbital phase. 
The error bars in the veiling values were obtained in the same manner as explained 
in Basri et al. (1991). 

We combined our radial-velocity data with those of Prato et al. (2002) to improve the 
orbital parameters of the binary system. We used the ORBIT program (Forveille 
et al. 1999) to find the best 
orbital solution to the data, which has a chi-square value of 192. 
The results of the best orbital fits to the data are given in Table 3 and are shown 
in Fig. 2. We note that the  radial velocity point of Prato et al. (2002) 
obtained at JD=2,452,311.2 significantly increased the error bars in the orbital elements and 
decreased the quality of the orbital fit. When this point was removed from the dataset, 
the chi-square value of the best orbital fit was 87, and the estimated 
masses of the components were changed significantly. In particular, the mass ratio q 
changed from 0.26 to 0.31. More radial velocity points for 
the secondary of UZ Tau E are clearly needed to improve the confidence of the mass estimates.

   \begin{figure*}
   \centering
   \includegraphics[width=\textwidth]{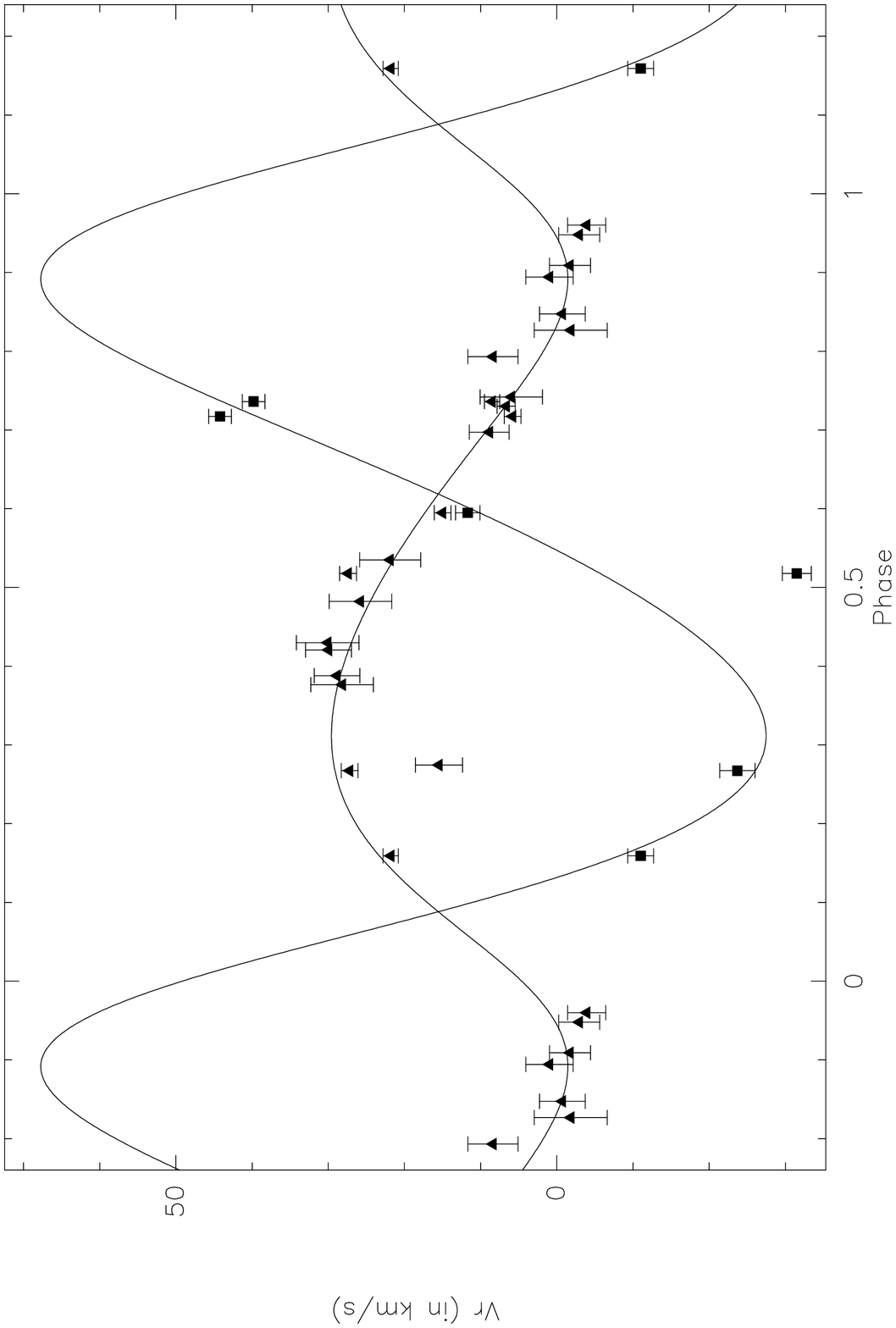}
   \caption{Phased radial velocity data for UZ Tau E. The triangles represent the primary star 
and the squares represent the secondary star (from Prato et al. 2002). 
The curves represent our best fit for these data.}
   \end{figure*}

\section{Discussion.}

The emission line spectrum, optical veiling, spectral energy distribution and millimeter 
emission of UZ Tau E are different facets of an accretion disk. 
Valenti et al. (1993) derived a mass accretion rate of 3$\times$10$^{-7}$ M$_\odot$ yr$^{-1}$ 
using blue low-resolution spectroscopy. In order to maintain such an accretion rate, circumbinary 
material must be flowing across the binary orbit. There is no evidence in the spectral energy 
distribution for a gap in the accretion disk, but the gap may be masked by a small 
amount of inflow material (Jensen et al. 1996). These authors infer a circumbinary disk 
mass of 0.06 M$_\odot$ from the millimeter continuum emission. 

The CTTS properties of UZ Tau E can be explained with the model developed by Artymowicz \& Lubow (1996) 
where accretion streams across the binary orbit allow large accretion rates and suppress 
orbital shrinkage by the flow of high-angular-momentum circumbinary material onto the stars. 
This model has also been invoked to explain the properties of the CTTS  
spectroscopic binary DQ Tau. A key prediction of this model is that 
accretion onto the stars in a binary system is modulated with the orbital phase. For high eccentricy 
binaries,  
most of the accretion takes place during periastron passage, but not for orbits 
with lower eccentricities. Observational support for this 
prediction was found in DQ Tau, where outbursts of emission lines and veiling were seen  
during periastron passages, although they were not present one third of the time 
(Basri et al. 1997; Mathieu et al. 1997). 
Moreover, in DQ Tau there is a permanent level of accretion going on at all 
orbital phases.  

   \begin{figure*}
   \centering
   \includegraphics[width=\textwidth]{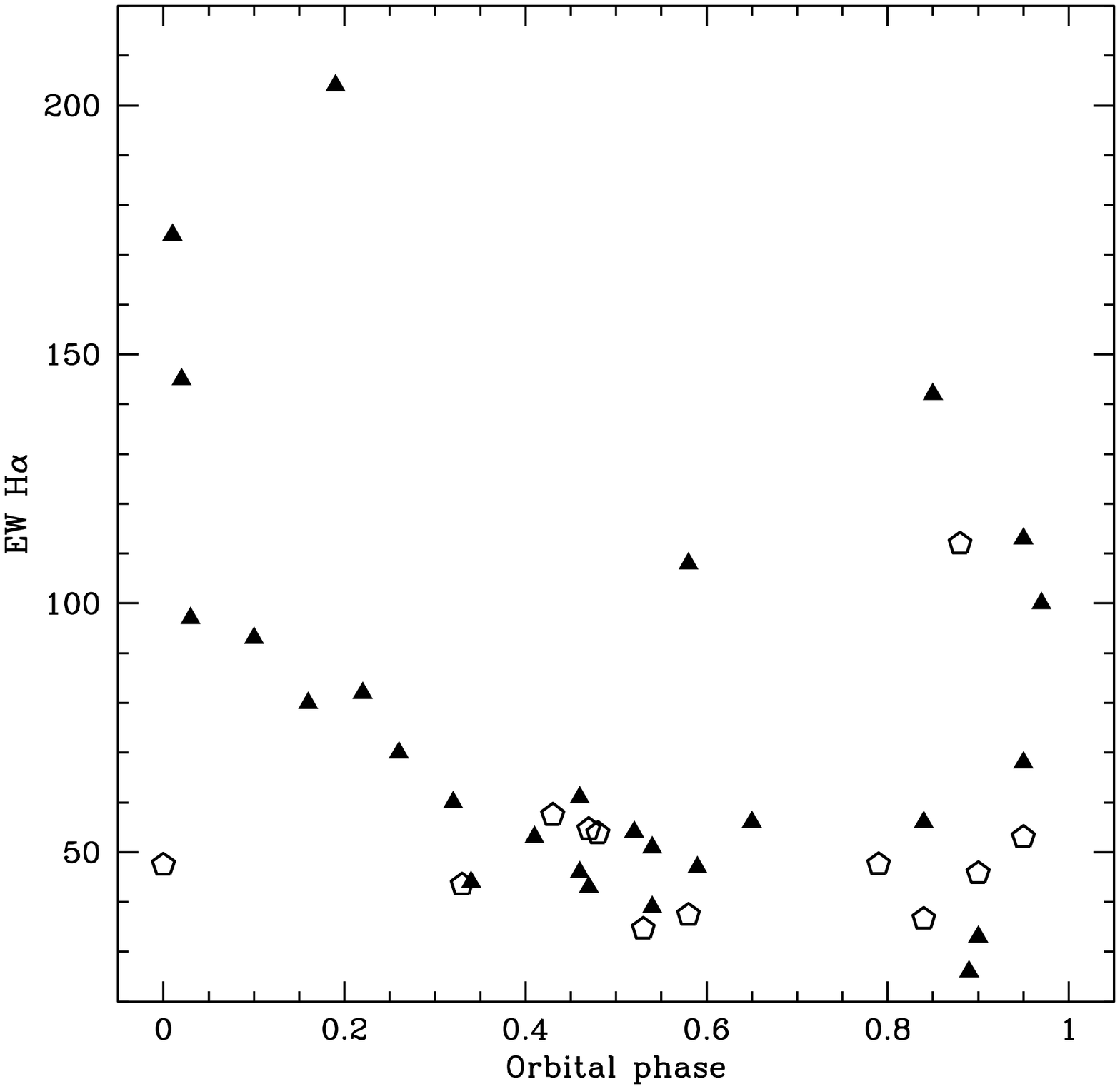}
   \caption{Phased  H$_\alpha$ emission  data for UZ Tau E (open pentagons) compared with 
DQ Tau (filled triangles, from Basri et al. 1997). 
      }
   \end{figure*}

   \begin{figure*}
   \centering
   \includegraphics[width=\textwidth]{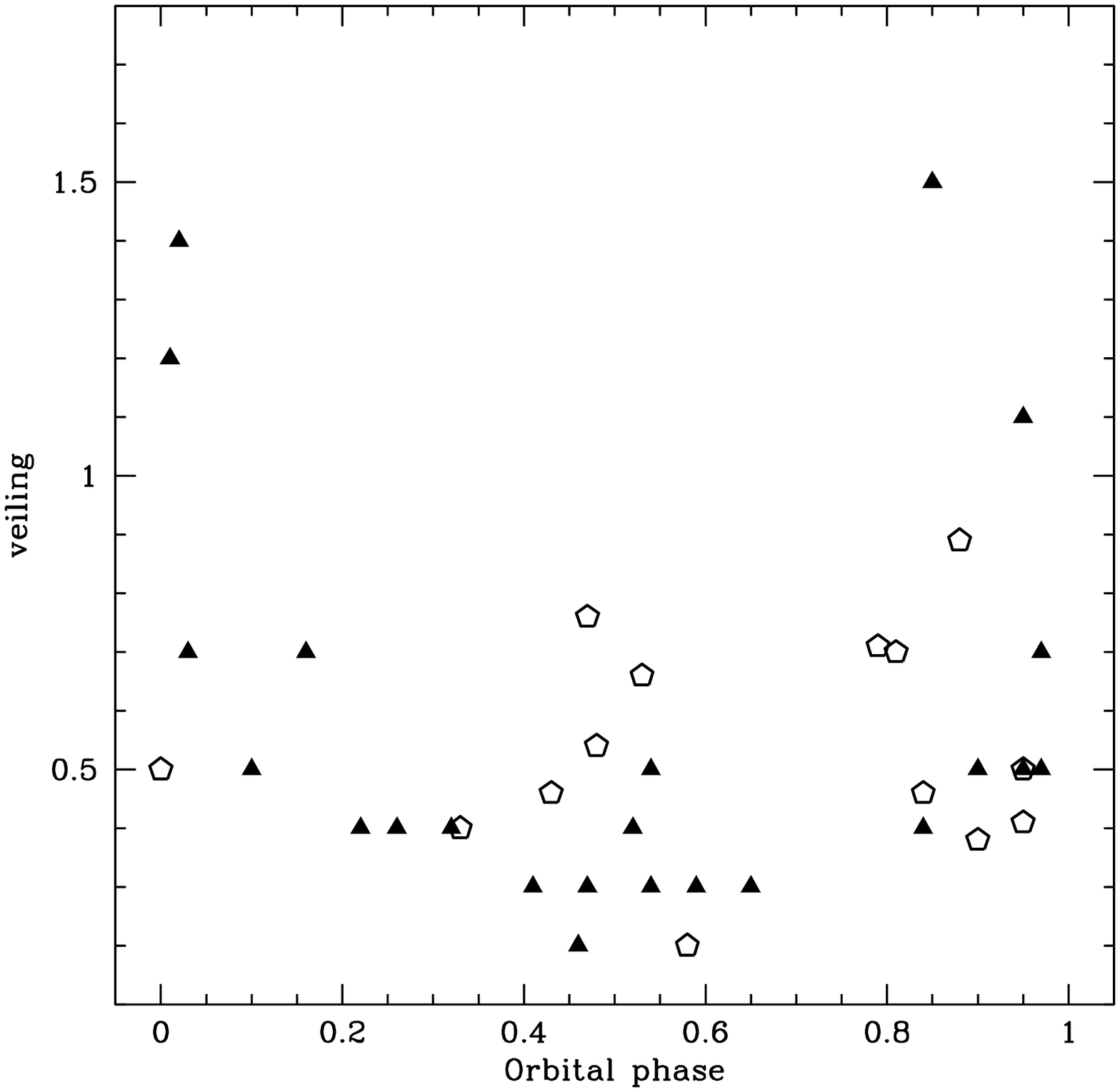}
   \caption{Phased  veiling measurements for UZ Tau E (open pentagons) compared with 
DQ Tau (filled triangles, data from Basri et al. 1997). 
      }
   \end{figure*}

It is interesting to compare UZ Tau E and DQ Tau. The orbital periods are similar (18.9 and 15.8 
days, respectively), but the eccentricities (0.14 and 0.55, respectively) and mass ratios 
(0.26 and 0.97, respectively) are very different. 
A comparison between the UZ Tau E and DQ Tau phased 
H$_\alpha$ emission datasets (without veiling corrections) is presented in Fig. 3. 
UZ Tau E has a level 
of permanent mass accretion comparable to DQ Tau as judged from the H$_\alpha$ equivalent width, 
which is on average 51.7 \AA , while in DQ Tau it is 79 \AA . However, 
we have not observed in UZ Tau E any outburst as strong 
as those seen in DQ Tau. We have one observation at JD=2,449,641.67 close to periastron passage 
(phase = 0.88) where we find the strongest H$_\alpha$ emission and largest veiling in our dataset. 
On the other hand, we do not witness any significant emission or veiling 
enhancement in the periastron passage from 
JD=2,450,039.49 to JD=2,450,042.66. 

In Fig. 4, we present a comparison of the dependence of veiling on orbital phase 
between UZ Tau E and DQ Tau. The veiling values for DQ Tau have been taken from Basri et al. (1997). 
The enhanced accretion events sometimes seen near periastron in DQ Tau are not present in 
our data of UZ Tau E. 
We conclude that the orbital modulation of accretion in UZ Tau E, if present, is not 
as clear as in DQ Tau. This could be due to the lower eccentricity of UZ Tau E or to 
the lower mass ratio of the two components, or to a combination of 
both parameters. Figure 2 of Artimowicz \& Lubow (1996) shows that their hydrodynamical models 
predict that accretion is much more enhanced at periastron passage in a binary with e=0.5 than 
in a binary with e=0.1. Hence, the difference in behaviour between these two CTTS binaries 
is qualitatively consistent with the models. 
 
Few other studies of the orbital dependence of accretion diagnostics in  
pre-main sequence spectroscopic binaries exist. 
Alencar et al. (2003) failed to find evidence for enhanced accretion near 
periastron in the pre-main sequence spectroscopic binary AK Sco which has an eccentricity of 
e=0.47 and an orbital period of 13.6 days. Stempels \& Gahm (2004) also did not find 
periastron activity in V4046 Sgr, a 2.4 day binary with a circular orbit.

Future studies of UZ Tau E are needed to improve the 
phase coverage of the accretion diagnostics, and to check the repeatability in its behaviour.
More CTTS spectroscopic 
binaries must be studied to determine observationally the role of the binary parameters 
on the time dependence of the accretion from the circumbinary disk. 
Enhanced periastron activity has clearly been seen in DQ Tau only, amongst 
the 4 pre-main sequence spectroscopic binaries observed so far. DQ Tau has the highest 
eccentricity among them, and hence it seems that enhanced accretion at periastron may 
occur only for extremely high eccentricities (e$>$0.5).

\begin{acknowledgements}
We thank Juan Manuel Alcal\'a for observing UZ Tau E with CASPEC and Gibor Basri for observing 
it at Lick and sharing the radial velocity measurements with us. 
We thank the observatory staff at La Palma observatory for assistance during 
the observations. 
\end{acknowledgements}

\clearpage




\clearpage

\begin{table*}
\caption{Observing log}
\begin{tabular}{lccccc}
\hline
\hline
UT date & JD (+2,400,000) & Instrument & 
Exp. (s) & Wav. range (nm) & FWHM (\AA )  \\
\hline
30 Nov 1988 & 47495.735 & Hamilton echelle & 2000 & 510-870 & 0.2    \\
16 Oct 1994 & 49641.672 & IDS/500mm/R600   & 420  & 593-674 & 2.4    \\  
6 Aug 1995  & 49936.723 & IDS/235mm/R1200  & 1200 & 653-673 & 1.6    \\
7 Aug 1995  & 49937.730 & IDS/235mm/R1200  & 1200 & 653-673 & 1.6    \\ 
8 Aug 1995  & 49938.727 & IDS/235mm/R1200  & 1200 & 653-673 & 1.6    \\
9 Aug 1995 & 49939.730 & IDS/235mm/R1200  & 1200 & 653-673 & 1.6    \\
12 Sep 1995 & 49972.730 & IDS/500mm/R1200  &  600 & 625-665 & 0.8    \\
19 Nov 1995 & 50039.495 & IDS/500mm/R1200  & 1000 & 635-674 & 0.8    \\
20 Nov 1995 & 50040.528 & IDS/500mm/R1200  &  900 & 635-674 & 0.8    \\
21 Nov 1995 & 50041.415 & IDS/500mm/R1200  &  900 & 635-674 & 0.8    \\
21 Nov 1995 & 50041.696 & IDS/500mm/R1200  &  900 & 635-674 & 0.8    \\
22 Nov 1995 & 50042.433 & IDS/500mm/R1200  &  900 & 635-674 & 0.8    \\
22 Nov 1995 & 50042.665 & IDS/500mm/R1200  &  900 & 635-674 & 0.8    \\
30 Nov 1995 & 50051.401 & IDS/500mm/R1200  & 1200 & 635-674 & 0.8    \\ 
12 Jan 1996 & 50095.449 & IDS/235mm/H1800  & 1250 & 624-677 & 1.1    \\
31 Jan 1996 & 50113.572 & CASPEC           & 1800 & 553-773 & 0.2    \\
3 March 1996 & 50145.653 & Hamilton echelle & 2000 & 510-870 & 0.2    \\
\hline
\end{tabular}
\end{table*}

\begin{table*}
\caption{Spectroscopic data for UZ Tau E}
\begin{tabular}{lcccc}
\hline
\hline
JD(+2,400,000) & RV (km s$^{-1}$) & EW(H$_\alpha$) (\AA) & v$_{\lambda}$ & Phase \\
\hline
47495.735 & 7.3$\pm$1.2  &          & 0.70$\pm$0.25\footnotemark[1]  & 0.81    \\
49641.672 & 4.9$\pm$4.8  &  112.1   & 0.89$\pm$0.20  & 0.88    \\
49936.723 & 34.9$\pm$4.1 &   57.5   & 0.46$\pm$0.20  & 0.43    \\
49937.730 & 36.8$\pm$4.1 &   53.8   & 0.54$\pm$0.20  & 0.48    \\
49938.727 & 32.5$\pm$4.1 &   34.6   & 0.66$\pm$0.20  & 0.53    \\
49939.730 & 28.6$\pm$4.0 &   37.4   & 0.20$\pm$0.20  & 0.58    \\
49972.730 & 22.2$\pm$3.1 &   43.5   & 0.40$\pm$0.20  & 0.33    \\
50039.495 & 15.1$\pm$3.3 &   36.6   & 0.46$\pm$0.15  & 0.84    \\
50040.528 &  6.0$\pm$3.0 &   45.8   & 0.38$\pm$0.20  & 0.90    \\
50041.415 &  7.7$\pm$3.1 &   53.0   & 0.41$\pm$0.15  & 0.95    \\
50041.696 &  5.0$\pm$2.7 &   52.6   & 0.50$\pm$0.15  & 0.95    \\
50042.433 &  3.8$\pm$2.7 &   47.9   & 0.50$\pm$0.20 & 0.00    \\
50042.665 &  2.8$\pm$2.5 &   46.7   & 0.50$\pm$0.20  & 0.00    \\
50051.401 & 36.7$\pm$3.0 &   54.6   & 0.76$\pm$0.20  & 0.47    \\
50095.449 & 12.7$\pm$4.1 &   47.6   & 0.71$\pm$0.15  & 0.79    \\
50113.572 & 15.6$\pm$2.6 &          &       & 0.74    \\
50145.653 & 29.5$\pm$3.0 &          &       & 0.44    \\
\hline
\end{tabular}

\thanks{\footnotemark[1] Veling value from Basri \& Batalha (1990).}
\end{table*}

\begin{table*}
\caption{Orbital parameters for UZ Tau E}
\begin{tabular}{lc}
\hline
\hline
Element & Value \\
\hline
Period & 18.979$\pm$0.007 days \\ 
Gamma  & 14.9$\pm$1.2 km s$^{-1}$ \\ 
K1     & 15.5$\pm$2.0 km s$^{-1}$ \\
K2     & 58.2$\pm$5.7 km s$^{-1}$ \\ 
ecc    & 0.14$\pm$0.05              \\
omega  & 217$^o$.9$\pm$30.1         \\  
T$_{\rm periastron}$      & 2,451,314.09$\pm$1.66 JD  \\ 
M1 sin$^3$(i) & 0.60$\pm$0.16 M$_\odot$  \\
M2 sin$^3$(i) & 0.16$\pm$0.04 M$_\odot$  \\
a1 sin(i)     & 0.02675 AU                 \\
a2 sin(i)     & 0.10056 AU                 \\
\hline
\end{tabular}
\end{table*}

\end{document}